\documentclass[10pt,conference]{IEEEtran}
\usepackage[utf8]{inputenc}

\usepackage{graphicx}
\usepackage{epstopdf}
\usepackage{float,url}
\usepackage[caption=false,font=footnotesize]{subfig}
\usepackage{amsmath,epsfig,psfrag,amssymb} 
\usepackage{amsfonts,graphics,graphicx}
\usepackage{algorithm,algorithmic}
\usepackage{textcomp}
\usepackage{xcolor}
\usepackage[nocompress]{cite}
\usepackage{multirow}

\newcommand{\ignore}[1]{}

\newtheorem{insight}{Insight}
\newtheorem{principle}{Principle}
\newtheorem{new-principle}{New Principle}

\begin{document}


\title{Analyzing Root Causes of Intrusion Detection False-Negatives: Methodology and Case Study}
\author{
\IEEEauthorblockN{Eric Ficke\IEEEauthorrefmark{1}, Kristin M. Schweitzer\IEEEauthorrefmark{2}, Raymond M. Bateman\IEEEauthorrefmark{2}, Shouhuai Xu\IEEEauthorrefmark{1}}
\IEEEauthorblockA{\IEEEauthorrefmark{1}Department of Computer Science, University of Texas at San Antonio\\
\IEEEauthorrefmark{2}U.S. Army Research Laboratory South - Cyber}
}
\maketitle


\begin{abstract}
Intrusion Detection Systems (IDSs) are a necessary cyber defense mechanism. Unfortunately, their capability has fallen behind that of attackers. This motivates us to improve our understanding of the root causes of their false-negatives. In this paper we make a first step towards the ultimate goal of drawing useful insights and principles that can guide the design of next-generation IDSs. Specifically, we propose a methodology for analyzing the root causes of IDS false-negatives and conduct a case study based on Snort and a real-world dataset of cyber attacks. The case study allows us to draw useful insights.
\end{abstract}

\section{Introduction}\label{intro}

Intrusion Detection Systems (IDSs) are an indispensable cyber defense tool. However, recent studies show that they have seen significant losses in detection effectiveness (see, e.g., \cite{divekar2018benchmarking,mireles2019metrics}). While it may be intuitive to speculate that attackers have been outmaneuvering IDS developers in this adversarial game of cat and mouse, as hinted in \cite{mireles2019metrics}, we need to precisely pin down the causes of such decreasing effectiveness. It is important then, to seek approaches to make IDSs as useful as possible.
This motivates us to investigate the root causes of intrusion detection false-negatives using modern cyber attack datasets with known ground truth.

\subsection{Contributions}\label{contributions}

In this paper we make the following contributions. First, we propose a methodology for systematically identifying and analyzing the root causes of IDSs' false-negatives. The methodology would be equally applicable to, or easily adapted to accommodate, any specific IDS and cyber attack datasets. 

Second, in order to show the usefulness of the methodology, we conduct a case study based on the widely-used Snort intrusion detection system against a modern cyber attack dataset, which is made available by the authors of \cite{shiravi-iscx}. This case study leads to the following findings:
(i) Some Snort rulesets have weak rules or miss certain attacks entirely, and it may be better to use more than one ruleset. 
(ii) Alerts do not always provide meaningful information, which may prevent human defenders from understanding the nature of the given attack.
(iii) Some rules do not follow Snort's design principles, which can provide a false sense of security without accurately detecting the intended attacks.
(iv) Some attacks do not utilize packet payloads in their implementation, so IDSs that rely on payload inspection may not be capable of detecting them.

\subsection{Related Work}\label{related-works}

By design, IDSs may use one of two approaches to identify cyber attacks: (i) comparing the traffic to known signatures of malicious use, or (ii) using machine learning to detect anomalous traffic \cite{axelsson-taxonomy}.
Signature-based detectors are well established and widely used to detect known attacks. However, this technique is limited because it requires expert knowledge of attack semantics to identify signatures and develop corresponding rules. This means that signature-based detection systems often suffer from the inability to detect 0-day attacks (e.g., \cite{song2008generalized}).
Anomaly detection (e.g., \cite{gao-wavelet,lu-rule-mining}) aims to go beyond this limitation of signature-based detection, but has encountered many obstacles preventing it from producing reliable results \cite{sommer2010outside}. As a result of these limitations, anomaly detection has not seen widespread deployment (although it continues to garner research effort in the hopes of achieving such viability). 
Some work has been done toward measuring the effectiveness of IDS \cite{ficke2018characterizing,mireles2019metrics}, but not the inherent limitations of IDSs.
Moreover, recent focus has shifted from the traditional packet-based view to a flow-based view for measuring IDS effectiveness \cite{sperotto-diss}.
This shift leads us to utilize flow-based view in our study.

To the best of our knowledge, this is the first paper that aims at systematically understanding and analyzing the root causes of intrusion detection false-negatives, with the ultimate goal of drawing new insights and principles to guide the design of the next generation IDSs. This is true despite the recent resurgence in studying IDSs (see, for example, \cite{ficke2018characterizing,mireles2019metrics,2018arXiv180603517H,Milenkoski:2015:ECI:2808687.2808691,Pendleton16,XuSTRAM2018ACMCSUR} and the references therein). In a broad sense, the present study falls into the broader field of cybersecurity data analytics \cite{XuBookChapterCD2019,XuCybersecurityDynamicsHotSoS2014,XuIEEECNS2014,XuCodaspy13-maliciousURL,XuIEEETIFS2013,XuIEEETIFS2015}.

\subsection{Paper Outline}
Section \ref{methods} describes the methodology for analyzing the root causes of intrusion detection false-negatives. 
Section \ref{sec:case-study} presents our case study.
Section \ref{discussion} discusses the limitations of the present study.
Section \ref{conclusion} concludes the present paper.

\section{Root Cause Analysis Methodology}\label{methods}

\begin{figure}[!htbp]
    \centering
    \includegraphics[width=.4\textwidth]{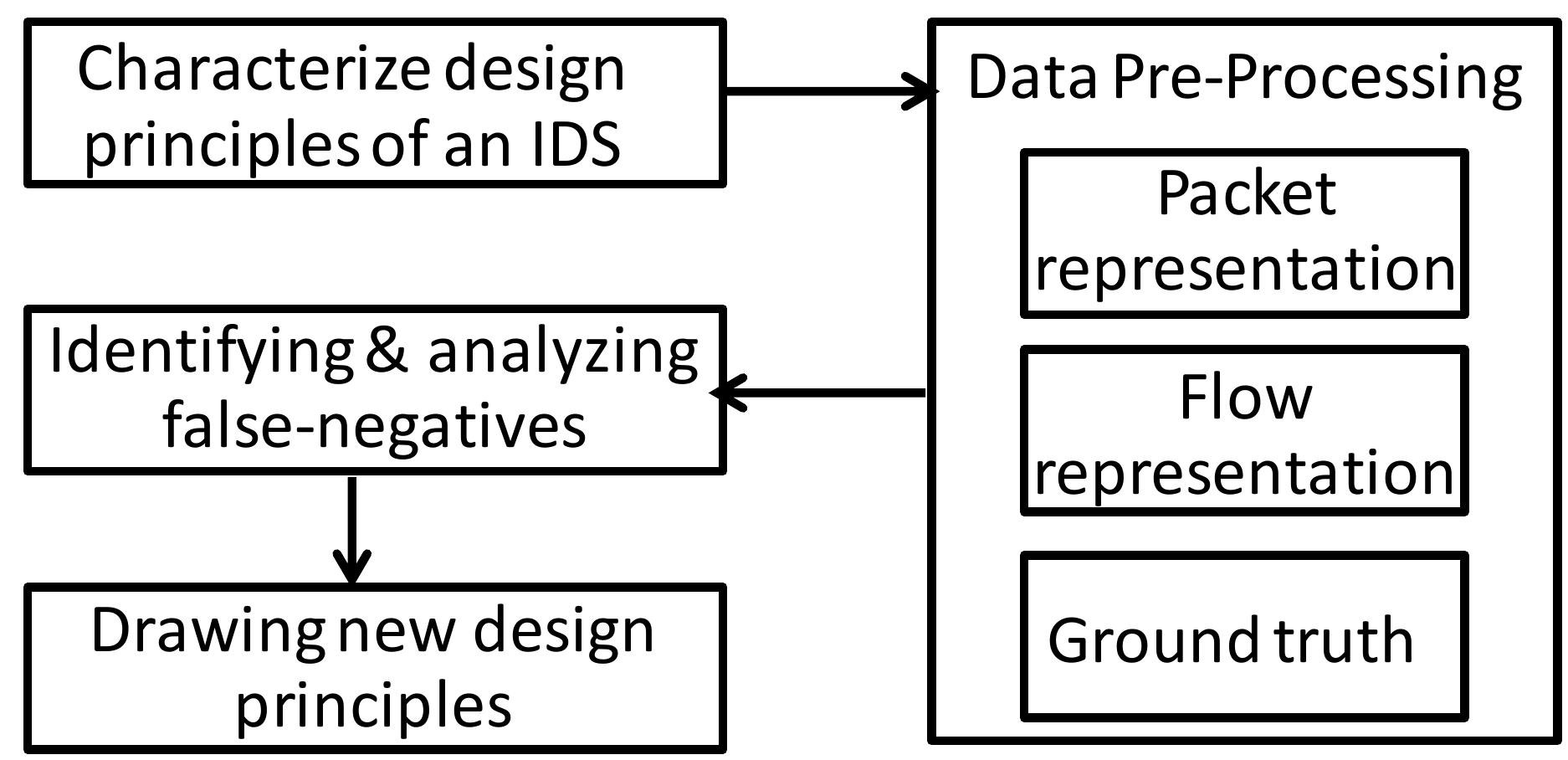}
    \caption{An overview of the root cause analysis methodology.}
    \label{fig:output-databases}
\end{figure}

We propose a {\em data-driven} approach to empirically identify false-negative instances of an IDS, and then analyze the root causes of the false-negative instances. 
This leads to the following methodology of four steps: (i) characterize the design principles of an IDS to understand its goals and define the scope; (ii) pre-process data with known ground truth;
(iii) identify and analyze false-negatives;
and (iv) draw new principles to to guide the design of the next generation IDSs.
These steps are highlighted in Figure \ref{fig:output-databases} and elaborated below.

\subsection{Characterizing Design Principles of an IDS}

Given an IDS, it is important to understand its design principles, and therefore its goals, because we cannot attribute a false-negative to the failure of an IDS if it is not designed to cope with such instances. That is, we need to understand what kinds of failures are indeed false-negative instances with respect to the goals of an IDS. For example, an IDS designed to cope with malicious uses may consider a ping sweep (i.e., a network scan using ICMP ping packets to identify online hosts) as malicious and label it as such, but an anomaly detector may observe similar traffic on a regular basis and therefore label the instance as normal. 

\subsection{Data Pre-Processing}

Suppose we are given a dataset of attacks (e.g., full packet capture with ground-truth tags), we propose creating two databases from the dataset: one for {\em flow-based} representation (i.e., treating each network flow as a processing unit), leading to a {\em flow-based database}; and the other for {\em packet-based} representation (i.e., treating each Internet Protocol or IP packet as a processing unit), leading to a {\em packet-based database}. 
This cross packet-flow analysis is important because newer IDSs are shifting toward flow-based alert generation \cite{sperotto-diss} and therefore cannot be compared to the packet-based alerts generated by legacy IDSs without using a cross packet-flow analysis. A network flow is typically specified by a tuple of 5 elements: {\em source IP address}, {\em destination IP address}, {\em source port number}, {\em destination port number}, and \textit{protocol}. Each flow has a start time (specifying when it starts) and a stop time (specifying when it is terminated).
We propose creating a mapping for each packet to its corresponding flow, in which each field of the 5-tuple must match and the packet's timestamp must fall between the corresponding flow's start and end times.

In order to create ground-truth tags at the packet or flow level, we can utilize the packet-flow mapping mentioned above for both the given ground truth and the IDS-generated alerts, either of which may be at the packet-level or flow-level.
Additionally, we prepare alerts by processing the network traffic according to the IDS's specifications for processing logged Packet CAPture (PCAP) files. The IDS must be configured to match the network architecture of the original network.

\subsection{Identifying and Analyzing False-Negative Instances}

Having pre-processed network traffic data as described above, we propose identifying false-negative instances as shown in the following steps.
First, we use the mapping between the packets and flows to match up the ground truth to the experimental results. With this, we can query the flow-based database to determine which attacks were identified correctly or not. For example, the query \texttt{flows.find(\{Tag:Attack,Alert:True\})}, would return all malicious flows which correctly resulted in IDS alerts. 

\begin{figure}[!htbp]
    \centering
    \includegraphics[width=\linewidth]{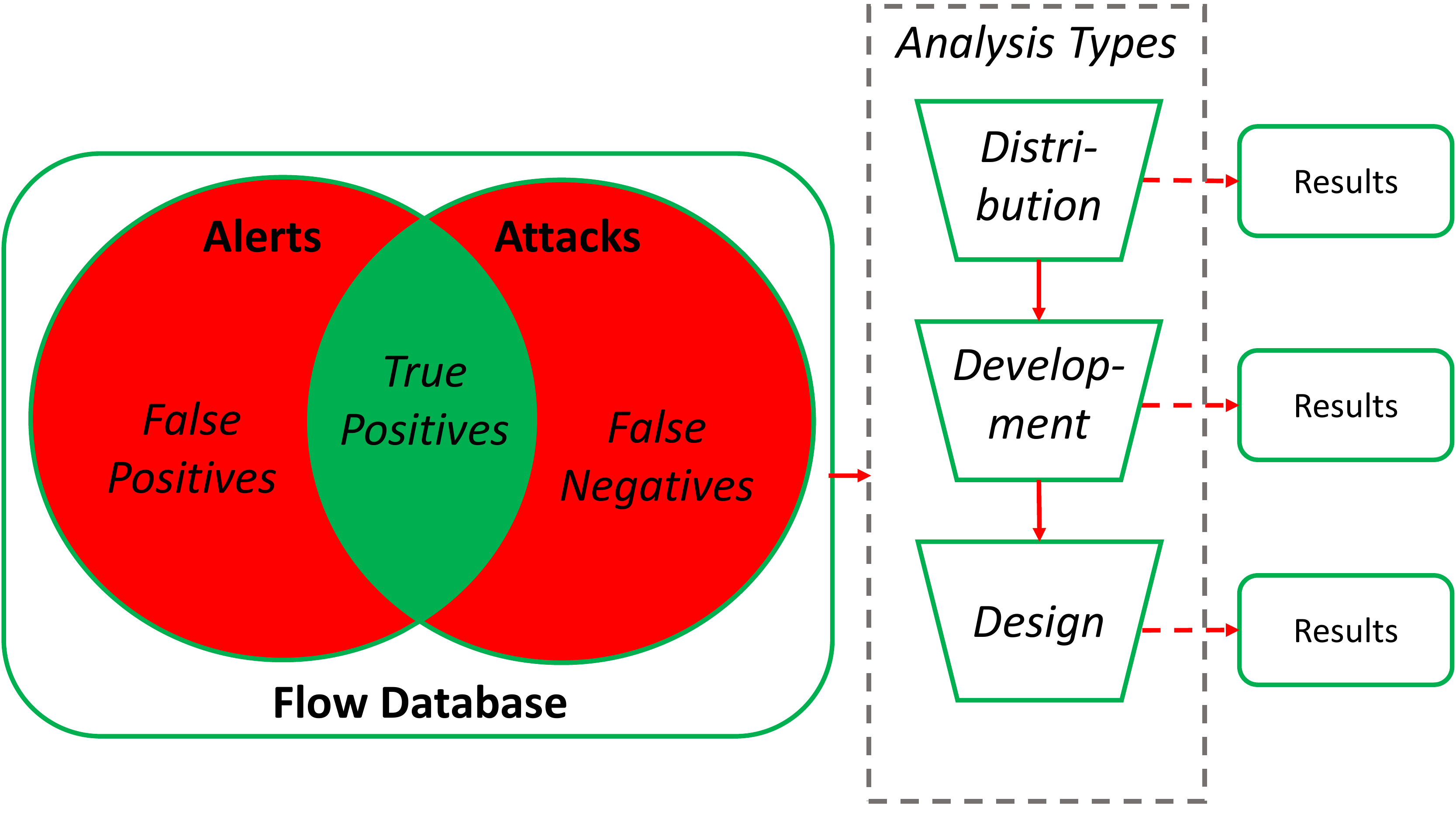}
    \caption{The detailed analysis procedure. Output from the flow database queries are scrutinized according to the IDS's development lifecycle, albeit in reverse.}
    \label{fig:methods-analysis}
\end{figure}

Second, we analyze the design of the IDS being tested, as shown in Figure \ref{fig:methods-analysis}, by considering flaws in the IDS's distribution system, the development process, and the design of the engine itself. 
The flaws that are identified are then synthesized into results.

\subsection{Drawing New Principles}

The previous step allows us to draw insights into the root causes of false-negatives and initiate an approach to reducing them. These insights will serve as a starting point for drawing guiding principles for designing future IDSs. These new principles may add to the existing set of principles that guided the design of the IDS in question, or may supersede some of them. 

\section{Case Study}\label{sec:case-study}
In this section we report our case study on applying the afore-presented methodology to identify false-negatives and analyze their root causes for Snort, which is a signature-based IDS. We choose Snort for the case study because it is one of the most popular IDSs and because it is an open-sourced system; the latter is important 
because we often need to analyze, for example, the rules or engines that are triggered. 
As a preliminary study, we use only the rulesets that are directly available from Snort at the time of installation, with no modification. 

\subsection{Characterizing Snort Design Principles} 

According to Snort's design principles (described in Section 3.9 of the Snort Manual \cite{snort}), the following should be considered when designing Snort rules for examining traffic.

\begin{enumerate}
\item[(i)] {\em Content matching:} For achieving a high performance in examining network traffic, the examination operation should match specific bytestrings so as to quickly rule out traffic that cannot match the rule in question. 
\item[(ii)] {\em Catching vulnerabilities, not exploits:} For assuring that rules cannot be evaded by the modification of exploit code, the examination operation should use rules that target vulnerabilities, meaning that these rules can be generalized so that they will identify any exploits that may be used against the vulnerability in question. This is important because there may be infinitely many exploits. 
\item[(iii)] {\em Considering protocol oddities:} ``Protocol oddities'' may include support for variable-length character encodings, format strings, etc, which may be leveraged by attackers to make their attack evade Snort. This principle hopes to assure that attackers cannot evade rules by making minor modifications to the protocols in question while obeying their specifications. Such rules should be able to understand the behavior of a target protocol so as to better determine whether a traffic is malicious or not.
\item[(iv)] {\em Checking discrete values before recursive ones:} In order to further improve Snort performance in examining network traffic, as with content matching, discrete values may allow one to quickly rule out unlikely-malicious network traffic and effectively reduce the time that is spent on processing a network packet. 
\item[(v)] {\em Optimizing for variable-length encodings:} The idea behind this principle is to improve Snort's performance by skipping over variable-length data if its content is not pertinent to an attack signature. By skipping over such data, Snort can reduce processing time by finding relevant fields faster. For example, a Remote Procedure Call (RPC) allows strings to be encoded by specifying a 4-byte length before the string, but fixed values may appear after such strings. In this case, reading the string length and skipping over an appropriate length in bytes may be faster than checking each byte in the string so that it must be completely parsed. 
\end{enumerate}
While several of these principles relate only to the runtime of Snort's processing, we take note of Principles (ii) and (iii) because they have important implications.
Specifically, if the rules that wrongly label network traffic do not follow these principles, we will attribute the inaccuracy to a failure of the rule developers in properly following Snort's design principles. In other words, the root cause of these false-negatives may be the poor design of Snort rules (i.e. attack signatures), rather than the poor enforcement of rules (i.e. distribution, processing or other aspects).

\subsection{Data Pre-Processing}
The dataset comes from a testbed network of user-generated attack traffic from the University of New Brunswick's Information Security Centre of Excellence (ISCX) \cite{shiravi-iscx}. The dataset contains mixed malicious and benign traffic from a realistic network architecture, including flow descriptors and ground-truth tags. More specifically, the dataset contains benign traffic generated from some statistical models of real traffic, as well as manually conducted attacks. We choose this dataset because it includes well documented and diverse attack scenarios and has been accepted by the research community for evaluating IDSs. Other datasets were considered, but these are more limited because they do not describe which attacks exist in the data, the data must be sanitized because of the presence of sensitive information, or they do not include complete packet captures. We use Python's MongoDB module, \textit{pymongo}, to manage a database for packets and a database for flows.  


The dataset contains nine distinct attacks \cite{shiravi-iscx},
which we categorize as follows, based on the goals of the IDS:
\begin{enumerate}
    \item[(I)] Vulnerability-leveraging attacks: These attacks explicitly target software vulnerabilities.     \begin{itemize}
        \item Adobe printf buffer overflow: This attack exploits a vulnerability that is caused by an input validation error in a PDF viewer application, which allows remote code execution.
        \item SMB stack overflow: This attack exploits a vulnerability that is an input validation error in Server Message Block (SMB) software, which allows remote code execution.
        \item SQL injection: This attack exploits a vulnerability that is an input validation error in the Structured Query Language (SQL) server, which allows remote code execution.
        \item Slowloris DoS attack: This Denial-of-Service (DoS) attack consumes the limited number of connections that a web server is able to establish at once, without producing any meaningful traffic.
    \end{itemize}
    \item[(II)] Auxiliary attacks: These behaviors are not inherently malicious, but are often leveraged by attackers.
    \begin{itemize}
        \item Reverse shell: An attack payload which causes the exploited host to establish a remote connection to the attacker's server so that the attack may issue further commands to their newly compromised victim.
        \item Nmap scan: A combination of ping and SYN packets sent to a broad range of IP addresses and ports in order to determine which hosts are online and listening on the network.
        \item IRC command \& control: Internet Relay Chat channels may used by attackers after exploiting a victim in order to issue further commands to compromised systems.
    \end{itemize}
    \item[(III)] Brute force attacks: These attacks do not target specific software vulnerabilities, but hope to overwhelm the target by sheer volume of attacks.
    \begin{itemize}
        \item DDoS (Distributed Denial-of-Service) attack: Bots create massive amounts of normal-looking traffic in order to slow down or freeze an individual system or network.
        \item SSH brute force login attempts: Bots attempt to gain unauthorized access to user accounts by repeatedly guessing passwords for known user accounts.
    \end{itemize}
\end{enumerate}

We focus on the category (I) attacks mentioned above because Snort's primary goal is to detect malicious uses. This can be further justified as follows. The category (II) activities mentioned above represent behaviors that are not inherently malicious (even though they may be leveraged in attacks); this justifies why we do not measure Snort's ability to detect attacks which fall in this category. Likewise, we exclude the attacks in category (III) for the following reasons. First, Brute force DDoS attacks are not caused by vulnerabilities in software; rather, they are the result of excessive volume of legitimate-looking traffic overwhelming the network and/or host resources \cite{mirkovic2004taxonomy}. Note that category (III) attacks do not include the Slowloris attack in category (I), which is a semantic DoS attack against a vulnerable protocol (i.e., the scale of traffic required for full effect is achievable with a single source machine, whereas a DDoS requires many more hosts to sufficiently affect the target). Second, the SSH brute force login attack targets the password generation practices of users; this is not a software vulnerability, so such attacks are not primary concerns according to Snort's design principles---instead, this attack can be better detected using an account-specific counter (as in practice). For these reasons, we only hold Snort accountable to the attacks in category (I).

\subsection{Identifying and Analyzing False-Negatives}\label{data-desc}

Now we analyze four instances of false-negatives encountered when using Snort to analyze the aforementioned dataset. In each case, we describe the attack, analyze the root cause of its false-negative, and draw some insight.

\subsubsection{A case of the Adobe printf buffer overflow attack}

\paragraph{The attack} 
This attack is executed by sending a malicious PDF file to a target users' email system as an attachment, which is a widely used social-engineering attack tactic. In the case described in the dataset,
a malicious PDF is sent to the user whose workstation holds IP address 192.168.1.105. The workstation uses the POP email protocol over port 110 to fetch the email from the mail server (192.168.5.122) \cite{shiravi-iscx}. Comparing this to the ground truth, we find that this interaction is not marked as malicious in the dataset's ground truth while it should be.
On the other hand, Snort generates several alerts corresponding to the traffic. Included in these are two unique signatures: ``{\tt [139:1:1] (spp\_sdf) SDF Combination Alert}'' (indicating the presence of sensitive data, such as email addresses) appears twice, and ``{\tt [129:12:1] Consecutive TCP small segments exceeding threshold}'' appears 8 times (indicating an anomaly detected by Snort's stream5 processor). However, it is not considered malicious for TCP segments to be unusually small, even in relatively high quantity. Neither of these alerts suggest the presence of a buffer overflow attack, indicating a false-negative.

\paragraph{Root cause analysis}
Snort's alerts do not describe, or even hint at, the presence of a buffer overflow attack; rather, these alerts give vague information from the sensitive data preprocessor (originally the ``Sensitive Data Filter'', or SDF) and small TCP segments (which has no semantic value to human defenders). Further exploration into an alternate ruleset (Emerging Threats \cite{emerging_threats_2019}) reveals that this attack is detectable by signature-based IDS. The rule is available as ``{\tt 2800385 - ETPRO WEB\_CLIENT Adobe Reader and Acrobat util.printf Stack Buffer Overflow}''. Since this is not in Snort's default ruleset, we conclude that it was not considered important for general use. Second, the fact that the attack is described in the paper describing the dataset \cite{shiravi-iscx} but is not tagged in the dataset raises concerns for the reliability of ground truth within the dataset.
From these observations, we draw two insights:

\begin{insight}
\label{insight-1}
Alternative rulesets must be considered because developers may not adopt rules designed by others into their own default set.
\end{insight}

\begin{insight}
\label{ins:atk}
Alerts must provide meaningful information that allows human defenders to understand the nature of an alert in question because it is not sufficient to simply report alerts without presenting useful or self-explaining information.
\end{insight}

\subsubsection{A case of the SMB stack overflow attack}

\paragraph{The attack} In the SMB stack overflow attack, a maliciously crafted packet is sent to the SMB server. The packet overflows a stack buffer on the target, allowing an attacker to execute arbitrary code. In this case, the attacker uses the vulnerability known as MS08-067 against the host at 192.168.2.113. Using our flow database, we access the relevant traffic marked as malicious. Comparing this to our Snort output, we find no alerts. Looking into the existing Snort rules, we surprisingly find several rules designed to detect (and marked explicitly for) MS 08-067. Unfortunately, none of these rules were triggered from the attack in the data. As such, this represents a false negative of the IDS.

\paragraph{Root cause analysis}
This false-negative is particularly surprising because the SMB stack overflow used in this instance (i.e. exploiting vulnerability MS08-067) is extremely well-known among the community and has been well-documented. The fact that Snort does not detect this attack with the default ruleset configuration is unexpected. In light of this, we consider another IDS (namely Suricata \cite{suricata}) to see if this problem is specific to Snort. After following the same processing procedure as before (using Suricata's default configuration and ruleset), Suricata indeed produced an alert corresponding to the attack. Specifically, Suricata generated alerts as ``{\tt [1:2008705:5] ET NETBIOS Microsoft Windows NETAPI Stack Overflow Inbound - MS08-067 (15)}''. Since this alert precisely describes the attack including the vulnerability which was targeted, it is considered sufficient for the sake of detection. Noting the difference between Snort's result and Suricata's, we draw:

\begin{insight}
\label{insight:3}
Even within a class of IDS (e.g. open-source, misuse detectors), different products have different detection capabilities. Thus, it is important to test each individually, rather than treating a class of IDSs as the sum of each product.
\end{insight}

\subsubsection{A case of the SQL injection attack}

\paragraph{The attack} In an SQL attack, the adversary submits a query with special characters to escape the unsanitized input field. Common versions of the attack include quotation marks, apostrophes and other special characters. Here, the attack is launched against the web server hosted at 192.168.5.123. We compare the ground truth with the alerts in the flow database. Of the 62 instances tagged as malicious, Snort created alerts for only 4. Specifically, they are labeled ``[129:12:1] Consecutive TCP small segments exceeding threshold''. As with the Adobe printf attack, this alert is not representative of the attack at hand, so we consider this a false negative.

\paragraph{Root cause analysis}
Some Snort alerts have been written in attempts to detect this attack, but they did not trigger in this instance. One example is ``{\tt [1:19439:8] SQL 1 = 1 - possible sql injection attempt}''. This rule looks for the string ``1=1'' in the packet, which is commonly used after the input string has been escaped. However, this string is specific to the exploit's implementation and not the vulnerability itself. As such, it clearly did not follow Snort's design principle, \textit{(ii) catching vulnerabilities, not exploits}. Following this observation, we draw the following insight:

\begin{insight}
\label{insight:4}
IDS rules must follow the principles established to guide their development.
As a first line of defense, this is especially important for open-source IDSs because attackers can analyze their rules when attempting to evade them without trial and error against live systems.
\end{insight}

\subsubsection{A case of the Slowloris DoS attack}

\paragraph{The attack}
The final attack we look at is the installation and execution of the Slowloris DoS software. In this attack, an attacker establishes numerous connections to a target server, until it has filled every socket available on the target, preventing it from accepting new connections from normal users. Slowloris is the name of a common software used for such an attack. In this case, the attackers utilize several of their previously compromised victims to amplify the efficacy of the attack, targeting the web server at 192.168.5.122. According to the tags included in the dataset, the compromised victims initiated 1969 connections to the target during their attack. In this version of Slowloris, we note that the packets sent do not contain payloads. Looking at the Snort output, we find several alerts showing the attacks labeled ``{\tt [139:1:1] (spp\_sdf) SDF Combination Alert}''. As with the Adobe printf attack, this alert is not representative of the actual attack in progress. As such, this is another false-negative.

\paragraph{Root cause analysis} 
Since Snort does not have any rules designed to detect attacks such as Slowloris, we look to other community-proposed rules that might be sufficient. One relevant rule, which is proposed in \cite{slowloris-stack-overflow_2019}, is based on a pattern identified from an instance of Slowloris (specifically, it identifies that the connection is kept alive by repeatedly sending arbitrary `X-a: ' header values). Unfortunately, the version of Slowloris in our data does not present the same pattern, meaning that even this rule would not catch the attack. Since this instance does not use payloads after the initial connection, it is important to find another means to detect the attack. In light of this, we draw the following insight:

\begin{insight}
\label{insight:5}
Signature-based IDSs are limited by their dependence on packet payload inspection, because some attacks do not utilize payloads.
\end{insight}

\subsection{Drawing New Principles}

Now we further abstract the insights drawn above into some new principles to guide the design of future IDSs. 
First, by summarizing and abstracting Insights \ref{insight-1} and \ref{insight:3}-\ref{insight:5}, we have:

\begin{principle}[clear specification]
The capabilities (and weaknesses) of an IDS should be specified, or even quantified, with respect to (i) the types of vulnerabilities or attacks it aims to defend against, (ii) the input it needs (e.g., header vs. payload information), and (iii) their guiding principles.
\end{principle}

Second, Insight \ref{ins:atk} highlights the importance of making alerts meaningful to defenders.
Previous study \cite{DBLP:conf/cyconus/MirelesCX16} has shown the importance of understanding the semantics of alerts, or more specifically the notion of {\em attack narratives}, as an effective approach to make sense of alerts. The notion of attack narratives aims to piece together information regarding several related attacks into a structured and human-understandable representation, which may be further put into the context of Mandiant’s Attack Life Cycle \cite{Mandiant} and/or Lockheed Martin’s Cyber Kill Chain \cite{CyberKillChainPaper2011}. That is, alert semantics are important for understanding the threat situation and responding to cyber attacks. Without meaningful alert semantics, human operators will be unable to quickly understand and respond to attacks. In light of this observation, 
we note that many attacks in the wild target known vulnerabilities, for which patches are often published at the time of their disclosure. Much of the analysis time for such attacks could be reduced if those vulnerable systems have already been patched. In other words, an intrusion detection system that is aware of the software states / versions running within an enterprise network may be able to predict whether an attack will be successful. This may provide a means to prune unnecessary alerts and reduce the interaction necessary between intrusion detection systems and human operators. This leads to:

\begin{principle}[IDS-defender interface]
IDSs must provide 
user-understandable alerts. One approach to achieving this is to make IDSs aware of the software stack postures (e.g., the presence of certain vulnerabilities). In this way, IDSs may achieve a higher capability in helping human defenders understand and respond to cyber attacks.
\end{principle}

Third, since intrusion detection systems often produce more alerts than what can be monitored by human defenders, it is important to present the most critical alerts to the human defenders. To this end, we propose defining standards by which the notion of {\em intrusion impact} may be quantified. With this concept, intrusion alerts may report on the worst-case or best-case impact estimates for a given attack. This leads to:

\begin{principle}[inherent alert prioritization]
IDSs themselves should prioritize alerts to defenders. Approaches to achieving this may include quantifying intrusion impact or describing attack narratives as mentioned above. This would enable prioritization of alert response.
\end{principle}

\section{Discussion}
\label{discussion}
The present study has the following limitations, which serve as motivations for future studies.
First, our methodology aims to associate packets to flows for standardizing a comparison between various IDSs. Several of the packets in the dataset used in the case study were not uniquely mapped to flows, so they were considered as belonging to all flows which matched the 5-tuple, with the timestamp requirement relaxed. 

Second, the case study is limited in scope to analyzing Snort, a malicious use detector. As such, it does not include anomaly detectors (e.g., IBM QRadar \cite{ibm-qradar-security-intelligence_2019} and Flowmon ADS \cite{flowmon-ads_2019}), which may not correspond directly to Snort.
Moreover, our study focuses on Snort and more specifically on its default ruleset, which means that we likely missed better detection capabilities of other popular rulesets (such as \cite{emerging_threats_2019} and \cite{suricata}). 

Third, the dataset we analyze has some limitations as well. In the course of analyzing the dataset, we notice the following issues with the data. (i) Some attacks which were described in the original paper that publishes the data did not appear to be tagged as such in the dataset. This made it difficult to precisely identify which alerts should be considered as relevant for some attacks, such as the Adobe printf attack. (ii) Some flows were tagged as malicious twice within the dataset. This could affect the measurement of true-positives. The present work is not affected by this issues, but those which measure detection rate may have been affected (e.g. \cite{ficke2018characterizing}). 

Fourth, much of our methodology is still manually operated, so applying it to various IDSs and datasets requires some time to conduct transitions between each stage, as well as the final querying of the database.

\section{Conclusion}
\label{conclusion}

We explained the importance of understanding the root causes of intrusion detection false-negatives. We introduced a methodology for this purpose and reported a case study on applying the methodology to analyze the root causes of Snort with respect to a real-world cyber attack dataset. We drew useful insights from this preliminary study. We hope that this study will inspire many future investigations into designing the next-generation intrusion detection systems.

\noindent{\bf Acknowledgements}. We thank Arash Habibi Lashkari for providing us the UNB ISCX 2012 dataset. This work was supported in part by ARL grant \#W911NF-17-2-0127 and NSF CREST Grant \#1736209.

\bibliographystyle{ieeetr}
\bibliography{eric}

\end{document}